\documentclass[conference]{IEEEtran}
\IEEEoverridecommandlockouts
\ifCLASSINFOpdf
\else
\fi
\usepackage[caption=false]{subfig}
\usepackage{cite}
\usepackage{upgreek}
\usepackage[pdftex]{graphicx}
\usepackage{cite}
\usepackage{xcolor,soul,framed} 
\colorlet{shadecolor}{yellow}
\usepackage{url}
\usepackage{array}
\usepackage{caption}
\usepackage{balance}
\usepackage{cite}
\usepackage{xcolor,soul,framed} 
\colorlet{shadecolor}{yellow}
\usepackage{color,soul}
\usepackage[pdftex]{graphicx}
\DeclareGraphicsExtensions{.pdf,.jpeg,.png}
\usepackage[cmex10]{amsmath}
\usepackage{algorithmic}
\usepackage{mdwmath}
\usepackage{mdwtab}
\usepackage{eqparbox}
\usepackage{url}
\usepackage{array}
\usepackage{fixltx2e}
\usepackage{dblfloatfix}
\usepackage[mathscr]{euscript}
\usepackage{amsfonts}
\usepackage{amsmath}
\usepackage{caption}
\usepackage{commath}
\usepackage{upgreek}
\usepackage[long]{optidef}
\usepackage{bm}
\usepackage{algorithm}

\usepackage[open,openlevel=2]{bookmark}
\usepackage{balance}
\usepackage{textgreek}
\usepackage{booktabs,chemformula}
\usepackage{array}
\usepackage{float}
\usepackage{amssymb}
\usepackage{graphicx}
\usepackage{bm}
\usepackage{cleveref}
\usepackage[thinlines]{easytable}

\makeatletter
\def\footnoterule{\kern-3\p@
  \hrule \@width 2in \kern 2.6\p@} 
\makeatother
\begin{document}
\IEEEoverridecommandlockouts

\title{Dynamic Selective Positioning for High-Precision Accuracy in 5G NR V2X Networks}
\vspace{-5 in}
\author{\IEEEauthorblockN{Abdurrahman Fouda, Ryan Keating
and Amitava Ghosh 
}
\IEEEauthorblockA{Nokia Bell Labs\\
abdurrahman.fouda@u.northwestern.edu, \{ryan.keating, amitava.ghosh\}@nokia-bell-labs.com}}

\maketitle
\begin{abstract}
The capability to achieve high-precision positioning accuracy has been considered as one of the most critical requirements for vehicle-to-everything (V2X) services in the fifth-generation (5G) cellular networks. The non-line-of-sight (NLOS) connectivity, coverage, reliability requirements, the minimum number of available anchors, and bandwidth limitations are among the main challenges to achieve high accuracy in V2X services. This work provides an overview of the potential solutions to provide the new radio (NR) V2X users (UEs) with high positioning accuracy in the future 3GPP releases. In particular, we propose a novel selective positioning solution to dynamically switch between different positioning technologies to improve the overall positioning accuracy in NR V2X services, taking into account the locations of V2X UEs and the accuracy of the collected measurements. Furthermore, we use high-fidelity system-level simulations to evaluate the performance gains of fusing the positioning measurements from different technologies in NR V2X services. Our numerical results show that the proposed hybridized schemes achieve a positioning error $\boldsymbol{\leq}$ 3 m with $\boldsymbol{\approx}$ 76\% availability compared to $\boldsymbol{\approx}$ 55\% availability when traditional positioning methods are used. The numerical results also reveal a potential gain of $\boldsymbol{\approx}$ 56\% after leveraging the road-side units (RSUs) to improve the tail of the UE’s positioning error distribution, i.e., worst-case scenarios, in NR V2X services. 
\end{abstract}

\IEEEpeerreviewmaketitle

\section{Introduction}\label{sec_intro}
Localization operations have been introduced in new radio (NR) vehicle-to-everything (V2X) services to estimate either the absolute coordinates of a vehicle or its relative position to the surrounding objects. It is of vital importance to provide NR V2X users (UEs) with stable and high-precision positioning accuracy because a small error may lead to a fatal accident. 5G Automotive Association (5GAA) has defined the positioning accuracy as a service level requirement (SLR) that ranges from a few centimeters to a couple of meters based on the nature of operation of each V2X use case~\cite{5gaaxn}. The 3rd Generation Partnership Project (3GPP) has introduced native positioning support in NR for handheld and indoor industrial use cases~\cite{npos}. For V2X use cases, sidelink in NR has been designed during Rel-16 without specific positioning support~\cite{usecases}. 

Generally, the proposed NR Rel-16 positioning techniques for handheld and indoor use cases are divided into (1) temporal methods, e.g., downlink time difference of arrival (DL-TDOA), (2) angular methods, e.g., downlink angle of departure (DL-AoD), and (3) hybrid schemes~\cite{overview}. It is expected that positioning enhancements to support positioning for V2X UEs, i.e., on-board units (OBU) in vehicles, may continue in Rel-17 and beyond. The potential positioning solutions for NR V2X services can be categorized into RAT-dependent, i.e., NR based, and RAT-independent techniques. In the former, V2X UEs use their cellular connections and the known locations of base stations (gNBs) for positioning using several techniques, e.g., DL-TDOA. In the later, V2X UEs exploit their satellite connections along with the known locations of the satellites for positioning using global navigation satellite system (GNSS) positioning techniques, e.g., code-based GNSS, PPP-RTK~\cite{esa}.

GNSS-based positioning has been considered as a promising solution in NR V2X services given the support of high-precision positioning, and the availability of many satellites. However, the limitation of the line-of-sight (LOS) connection in many scenarios, e.g., dense urban environments, tunnels and underground parking, is among the main challenges to achieve high accuracy~\cite{5gaaxn}. On the other hand, perfect network synchronization, support of high mobility, reliability requirements, the minimum available number of anchors, and bandwidth limitation are among the main challenges to provide high-precision positioning using RAT-dependent techniques~\cite{oppr}. 

The fusion of positioning measurements from different technologies has emerged as a potential solution to improve the positioning accuracy in V2X services (see, e.g., Section 7.2.3 in~\cite{nrpos_855} and references therein). However, the mutual dependence between the deployment environments and the accuracy of the collected measurements is one of the main challenges to achieve high accuracy using these hybridized techniques~\cite{sony}. Essentially, fusing two groups of accurate and poor measurements will not lead to an improvement in the overall accuracy. Hence, a new selection method is needed to determine when to have a near-optimal bias toward one technology over the other based on the locations of V2X UEs and the accuracy of the collected measurements.

A least squares-based method for hybridization of GNSS and RAT-dependent measurements was presented in~\cite{esa2},~\cite{huawei}. However, the authors did not discuss how the optimal weights can be determined to have a specific bias for one positioning technology over the other. To the best of our knowledge, none of the prior art discussed the selection between different positioning technologies to improve the positioning accuracy in NR V2X services. In this paper, we propose a novel selection method; namely, selective positioning based on neighboring ToA variance (SPNTV) to dynamically switch between the RAT-dependent and GNSS positioning methods. 

Our numerical analysis shows that the proposed SPNTV achieves a positioning error $\leq$ 3 m with $\approx $ 76\% availability compared to $\approx $ 51\% and 58\% availability when GNSS and DL-TDOA measurements are solely used, respectively. We also evaluate the performance of a weighted MAP-based fusion method and show the advantage of using the proposed SPNTV over the fusion method in several V2X use cases. Further, we present enhanced versions of the above hybridized positioning methods that allow V2X UEs to dynamically switch between positioning schemes (not only different technologies) based on their locations. Finally, we demonstrate the performance gains of utilizing the road-side units (RSUs) to improve the tail of UE’s positioning error distribution, in NR V2X services. 


\section{Network Architecture and Simulated Environment}\label{sec_sysmodel}
\begin{figure}
  \begin{center}
  \includegraphics[width=8cm,height=8cm,,keepaspectratio]{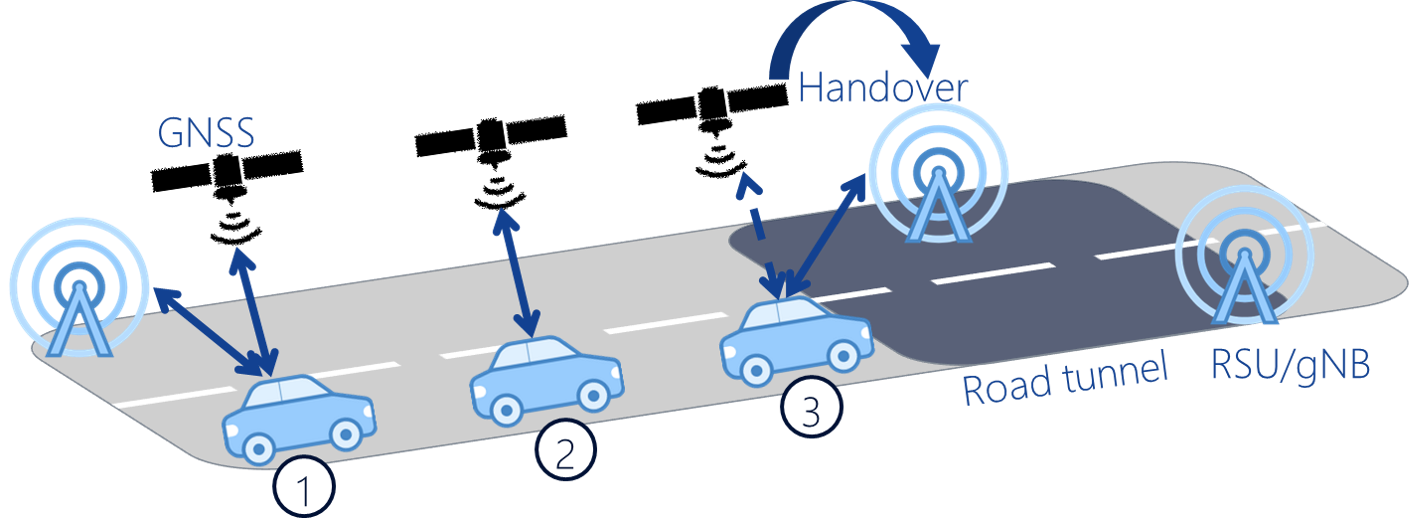}
  \caption{Different positioning modes for NR V2X services .}\label{fig_sysmod}
  \vspace{-.25in}
  \end{center}
\end{figure}

In this paper, we use a two-tier cellular network to analyze the positioning performance of two NR V2X use cases. In the first use case, we assume that V2X UEs are distributed randomly at uniform in a six-lane urban road. In the second one, we assume that a tunnel section is added to the urban road. In that, V2X UEs are divided into two equal groups, one inside the tunnel and another outside the tunnel. As shown in Fig.~\ref{fig_sysmod}, a V2X UE can use several modes for positioning estimation. In mode 1, V2X UE fuses the GNSS and cellular measurements to perform positioning. Mode 2 represents a single technology positioning method, in which, V2X UE uses only GNSS measurements to perform positioning. Similarly, V2X UE uses only cellular measurements to perform positioning in mode 3.  

\subsection{NR TDOA-based Positioning}\label{sec_tdoa}

We focus on DL-TDOA (as a cellular positioning method), but the proposed algorithms are not limited to only this method. A new reference signal for positioning (DL PRS) was introduced in Rel-16, which we use in this work for positioning estimation~\cite{overview}. In particular, V2X UEs measure the reference signal time difference (RSTD) between DL PRS from different transmission points, i.e., gNBs and RSUs, to perform positioning~\cite{oppr}. The V2X UEs use the legacy LTE positioning protocol (LPP) to exchange location information with the location server. A trilateral estimation algorithm is then used by the location server to estimate the positions of V2X UEs given the known locations of gNBs, RSUs, and the RSTD measurements. It is worth mentioning that we assume that DL-TDOA is supported in both NR V2X UE-based (i.e., UEs perform positioning estimation) and network-based mode. In general, the algorithms and solutions we describe in this paper apply to both UE-based and network-based positioning.  

\subsection{GNSS-based Positioning}\label{sec_gnss}
GNSS positioning relies on receiving signals from multiple satellites at V2X UEs and making pseudo-range measurements from these received signals. Based on the known locations of the satellites and the pseudo-range measurements, the UEs can estimate their locations. Let
$\mathscr{S}=\left\{1,\dots,S\right\}$, $\mathscr{U}=\left\{1,\dots,U\right\}$
denote the sets of GNSS satellites and V2X UEs, respectively where, e.g., the cardinality of $\mathscr{S}$ is $\abs{\mathscr{S}}=S$. The GNSS pseudo-range measurements of $u^{\text{th}}$ V2X UE is given by~\cite{esa2}:

\begin{equation}\label{pseudorange}
{\tilde{\mathbf{d}}}_{u}= \mathbf{d}_{u}+\mathbf{e}_{u} \forall u\in\mathscr{U}, 	    
\end{equation}
where $\mathbf{d}_u=\left({d_{}}_{u,s}:s\in\mathscr{S}\right)$ with ${d_{}}_{u,s}$ represents the distance between $u^{\text{th}}$ V2X UE and $s^{\text{th}}$ satellite and is given by $d_{u,s}=\norm{\mathbf{p}_u-\mathbf{p}_s}$,
where ${\mathbf{p}_{}}_u=\left[x_u,y_u,z_u\right]^\top$ and ${\mathbf{p}_{}}_s=\left[x_s,y_s,z_s\right]^\top$ are the positions of $u^{\text{th}}$ V2X UE and $s^{\text{th}}$ satellite, respectively. Similarly, pseudo-range error between $u^{\text{th}}$ V2X UE and GNSS satellites is given by $\mathbf{e}_u=\left({e_{}}_{u,s}:s\in\mathscr{S}\right)$, where we model ${e_{}}_{u,s}$ as a Gaussian distribution with $\mathcal{N}\left(0,\mathbf{1}_M^\top\sigma^{2}\right)$, ${\mathbf{\sigma}_{}}^2=\left[\sigma_1^2,\dots,\sigma_M^2\right]^\top$ represents the vector of user equivalent range error (UERE) variances and $\mathbf{1}_{M}$ denotes the $M$-dimensional all-ones vector. We define UERE variance error based on the LOS status between $u^{\text{th}}$ V2X UE and $s^{\text{th}}$ satellite $\rho_{u,s}$ as:
\begin{equation}
  \mathbf{\sigma} =
    \begin{cases}
      \left[\sigma_{o},\sigma_{i},\sigma_{t},\sigma_{n}\right] & \text{$\rho_{u,s}=0$,}\\
      \left[\sigma_{o},\sigma_{i},\sigma_{t},\sigma_{n},\sigma_{m}\right] & \text{$\rho_{u,s}=1$,}\\
      
    \end{cases}       
\end{equation}
where $\sigma_o$,$\sigma_i$,$\sigma_t$,$\sigma_n$, and $\sigma_m$ denotes the variance of the GNSS clock, residual ionosphere, troposphere, additive white Gaussian receiver noise, and urban multipath error, respectively.

\section{Proposed Hybridized Positioning Techniques}\label{sec_hymethods}
In this paper, we jointly utilize the positioning measurements collected from different technologies to improve the achievable accuracy in NR V2X services. In particular, we propose two hybrid positioning schemes; namely, SPNTV and weighted-MAP based fusion. In the former, V2X UEs utilize the ToA measurements to dynamically switch between different technologies based on the estimated accuracy of the collected measurements. In the latter, V2X UEs fuse the positioning measurements collected from several technologies to perform positioning. We assume a V2X UE can perform both GNSS and DL-TDOA positioning in a UE-based mode. 

\subsection{Proposed SPNTV Algorithm }\label{sec_spntv}
In this section, we discuss how the estimated ToA measurements can be used to define the selection criteria of the proposed novel SPNTV scheme. Essentially, V2X UEs use the received DL PRS from different cells to calculate the estimated ToA. We use Monte Carlo system-level simulations to study the performance characteristics of the estimated ToA measurements from the serving cell and the closest $\beta$ neighbors. In doing so, we consider modified urban macro (UMa) and micro (UMi) scenarios, in which, V2X UEs are randomly distributed over $l$ lanes in a two-tier cellular network following the same mobility modeling in~\cite{36885}.

We consider two use cases for the road environment in V2X services; urban road, and urban road with a tunnel, in which, V2X UEs suffer from penetration loss and NLOS blockage while being inside the tunnel. The channel realizations and V2X UE locations are randomly updated using the Monte Carlo simulations (see Section IV). To this end, V2X UEs are classified into two classes based on the estimated positioning error using the DL-TDOA technique. In that, class $\mathbb{A}$ represents V2X UEs  with an estimated positioning accuracy $\leq \epsilon$, where $\epsilon$ denotes the absolute positioning accuracy target of at least 95\% of V2X UEs according to the NR V2X SLRs in~\cite{5gaaxn},~\cite{slr}. On the other hand, the V2X UEs with an estimated positioning accuracy $>\epsilon$ are represented by class $\mathbb{B}$. 

\begin{figure}
  \begin{center}
  \includegraphics[width=8.5cm,height=8.5cm,,keepaspectratio]{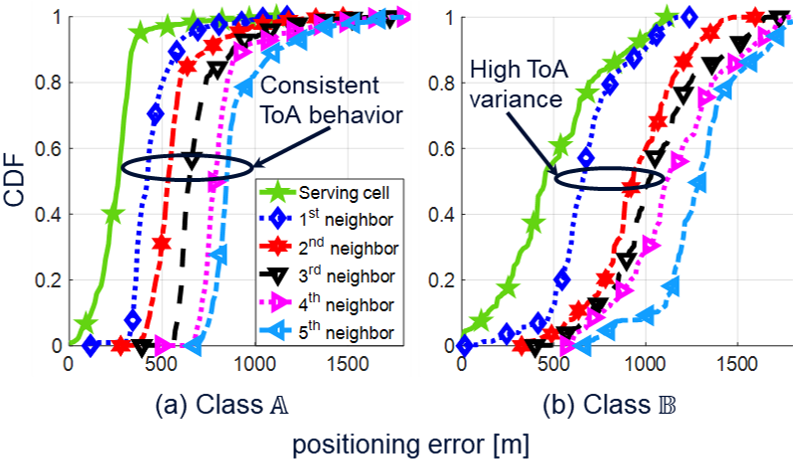}
  \caption{TOA measurements of V2X UEs.}\label{fig_toa}
  \vspace{-.25in}
  \end{center}
\end{figure}

Fig.~\ref{fig_toa}-a depicts the estimated ToA measurements with the serving cell and the first $\beta$ arriving neighbors to the class $\mathbb{A}$ of V2X UEs, where $\beta=5$. It shows that the ToA measurements decrease as the V2X UE becomes closer to its neighbor cell. Similarly, Fig.~\ref{fig_toa}-b shows the same trend with the class $\mathbb{B}$ of UEs. Figs.~\ref{fig_toa}-a and~\ref{fig_toa}-b reveal that the ToA measurements of the first arriving pair of neighbors have a high variance when V2X UEs suffer from a positioning error $>\epsilon$. It is worth noting that a consistent high variance between the ToA measurements of the closest pair of neighbors implies a long-term poor DL-TDOA positioning performance. 

We utilize the above findings and introduce a new parameter $\zeta=t_1-t_2$, where $t_i$  is the ToA of the $i^{\text{th}}$ arriving neighbor PRS in time, and $\zeta$ denotes the time difference between the ToA measurements of the closest pair of neighbors to a V2X UE. In that, V2X UE uses the first arriving pair of ToA as the measurements of the closest pair of neighbors. V2X UEs use the dependency between $\zeta$ (i.e., the variance of neighbor ToA measurements) and the estimated accuracy of the DL-TDOA to decide whether to use the cellular connection for positioning or to switch to another technology. We implement such dependency in a binary fashion, in which, V2X UE  only decides to use DL-TDOA technique for positioning if $\zeta<\eta$, where $\eta$ denotes a triggering event to switch between different positioning technologies. We initially introduce $\eta$ to be the $50^{\text{th}}$ percentile of $\boldsymbol{\zeta}_{\mathbb{A}}=\left(\zeta_{a}:a\,{\in}\,\mathbb{A}\right)$, where $\zeta_a$ represents the difference between the ToA measurements of the closest pair of neighbors to a class $\mathbb{A}$ V2X UE. In this paper, we use the GNSS as an alternative technology that V2X UEs can switch to, based on the estimated measurements of $\boldsymbol{\zeta}_{\mathbb{A}}$. The proposed SPNTV scheme is summarized in Algorithm~\ref{algo_1}. 

\begin{algorithm}
\caption{Dynamic selection between GNSS and DL-TDOA using the proposed SPNTV scheme.}\label{algo_1}
\begin{algorithmic}[1]  
	\STATE\textbf{Inputs} base station and GNSS locations.
	\STATE\textbf{Initialization} $\eta=50^\text{th}\%$-tile of $\zeta_{\mathbb{A}}$, where $\zeta_{\mathbb{A}}$ is computed via simulation campaigns.
	\STATE\textit{Start}
	\STATE V2X UE measures DL PRS from both serving and neighboring cells and calculates ToA. 
	\STATE Calculate $\zeta = t_1-t_2$.  
	\IF {$\zeta\ge\eta$}
	\STATE 	V2X UE switches the GNSS receiver on and 	obtains a new GNSS positioning fix.
	\STATE 	Final V2X UE location is determined using only GNSS.
    \ELSE
	\STATE	V2X UE switches the GNSS receiver off.
	\STATE Final location is determined using only DL-TDOA.
	\ENDIF
	\STATE Refine selection parameter $\eta$ and \textbf{goto} \textit{Start}.
\end{algorithmic}
\end{algorithm}

It is worth mentioning that the predefined selection threshold $\eta$ may be dependent on the scenario, e.g., small cell, and inter-site distance. Also, $\eta$ can be either computed via simulation campaigns (see Section~\ref{sec_results} for an example) or determined by test V2X UEs in the field. Alternatively, a reference V2X UE can learn $\eta$ using machine learning over time. For example, by comparing relative DL-TDOA or GNSS positioning accuracy with a reference location and determining $\eta$ which leads to the highest positioning accuracy. The V2X UE may also share the determined $\eta$ with nearby V2X UEs via sidelink to further refine the selection. The entire procedure may be repeated periodically to update the selection of technology, and the periodicity may be based on either the velocity of V2X UE, e.g., update faster for highly mobile UE, or the absolute value of $\zeta$, e.g., for values close to $\eta$ update faster. It is also worth noting that while the method is described for DL-TDOA it could also apply to multi-cell round trip time or other future timing-based positioning methods in NR.

\subsection{Weighted MAP-based fusion Algorithm}\label{sec_wmap}
In this section, we present the implementation of a weighted version of the maximum a posteriori (MAP) estimator to fuse both DL-TDOA and GNSS measurements. Compared with SPNTV, the MAP fusion algorithm uses both technologies for positioning, instead of selecting just a single technology. The MAP estimator initially presented in~\cite{blade} for DL-TDOA measurements is used as a baseline. Given the known locations of gNBs and satellites, V2X UEs utilize the MAP estimator for positioning. It is worth noting that the same estimator is used for positioning estimation using either the DL-TDOA or GNSS measurements only (see Sections~\ref{sec_tdoa},~\ref{sec_gnss}). Let $\mathscr{G}=\left\{1,\dots,G\right\}$ and $\mathscr{C}$ denote the sets of gNBs and overall anchors, respectively, where $\mathscr{C}=\mathscr{S}\cup\mathscr{G}$ and $C=S+G$. The vector of fused measurements at $u^\text{th}$ V2X UE is given by $\mathbf{m}_u=\left(m_{u,c}:c\,\in\,\mathscr{C}\right)$, in which, $m_{u,c}=\Tilde{d}_{u,c}$ if $c\,\in\,\mathscr{S}$ and $m_{u,c}=\Tilde{t}_{u,c}$ if $c\,\in\,\mathscr{G}$ where $\Tilde{t}_{u,c}$ is the estimated ToA measurement between $u^\text{th}$ V2X UE and $c^\text{th}$ gNB. The MAP estimator is given by: 
\begin{equation}
    \begin{split}
        &\left(\Tilde{\mathbf{p}}_u,\Tilde{\tau}\right)=\arg\max_{\mathbf{p}_u,\tau} \prod_{c\,\in\,\mathscr{C}} w_{u,c}+\frac{1}{\left(\norm{{\mathbf{p}}_{u}-{\mathbf{p}}_c}+\epsilon\right)\sqrt{2\pi\sigma_{c}^2}}\\ &\exp{\left[\frac{-1}{2\sigma_c^2}\left(m_{c,u}-\frac{\norm{{\mathbf{p}_{u}-{\mathbf{p}}_{c}}}}{c}-\tau-\mu_{c}\right)\right]}, \forall\,u\in\mathscr{U},
    \end{split}
\end{equation}
where $\Tilde{p}_u$,$\Tilde{\tau}$, $\epsilon$ and $c$ represent the estimated transmit time (unique for all anchors), the estimated position of V2X UE, an offset to prevent numerical instability, and the speed of light, respectively. $w_{u,a}$ denotes the bias toward GNSS or DL-TDOA measurements. Finally, the measurement estimation error of $c^\text{th}$ anchor follows a Gaussian distribution with $\mathcal{N}(\mu_{c},\sigma_c^2)$.  

\section{Numerical Analysis}\label{sec_results}
In this section we investigate the achievable accuracy of the proposed hybridized schemes for a Rel-16 PRS at frequency range 1 (FR1) with bandwidth of 100 MHz. It should be noted that the proposed positioning methods also apply to Rel-17 and beyond where the accuracy could be greatly improved due to PRS enhancements or wider BW, e.g., using carrier aggregation and millimeter-wave (mmWave) frequency band. For single technology-based positioning, we use the DL-TDOA and GNSS measurements given the known locations of gNBs and satellites, respectively. Furthermore, we fuse both measurements and use a weighted MAP-based fusion algorithm for positioning (see Section~\ref{sec_sysmodel}). We compare the performance of the proposed SPNTV method with each one of the previous positioning techniques in two V2X UE dropping use cases; namely, urban road and tunnel use cases. 

\begin{table}
\renewcommand{\arraystretch}{1.3}
  \centering
  \caption{Simulation parameters.}\label{tab_simpara}
        \begin{tabular}{l l}
        \hline
            \textbf{Parameter}  & \textbf{Value}\\
        \hline
            Layout & 3GPP UMi, 57 cells, ISD = 200m~\cite{huawei}.\\
        \hline
            Channel model  & 3GPP cellular, NTN SCM for 0.5-100 GHz\\
            & ~\cite{38811},~\cite{38901}.\\
        \hline
           GNSS constellation   & 11 NGSO satellites at 1500km [15].\\
        \hline
            Carrier bandwidth  & Cellular: 100 MHz at 4 GHz.\\
            &GNSS: 24.5 MHz at 1.5GHz.\\
        \hline
            NR numerology   & 30 KHz subcarrier spacing.\\
        \hline
            Reference signal   & PRS, comb-6, comb-offset=0.\\
        \hline
            Mobility settings & Option A, speed=140 km/h, 6 lanes,\\
            & density = 600 V2X UEs/km~\cite{36885}.\\
        \hline
            Antenna geometry  & \shortstack{$2\times2$, SU-MIMO, MMSE receiver.}\\
        \hline
            Penetration loss  & \shortstack{20 dB (in tunnel use case).}\\
        \hline
            SPNTV settings  & $\beta=5$, $\epsilon=10$m,\\
            &$\zeta_{\mathbb{A}}= {16,55,90,155,220}$ ns.\\
        \hline
            MAP setting  & $w_{u,s}=\{3e-7,3e-8\}$, $\mu_a= 0$, \\ 
            &$w_{u,g}= \{1e-3,3e-4\}$, $\sigma_a= 50$.\\
            \hline
        %
        \end{tabular}
        \vspace{-.15in}
\end{table}

\subsection{Urban road use case}\label{case1}
\begin{figure}
  \begin{center}
  \includegraphics[width=6.5cm,height=6.5cm,,keepaspectratio]{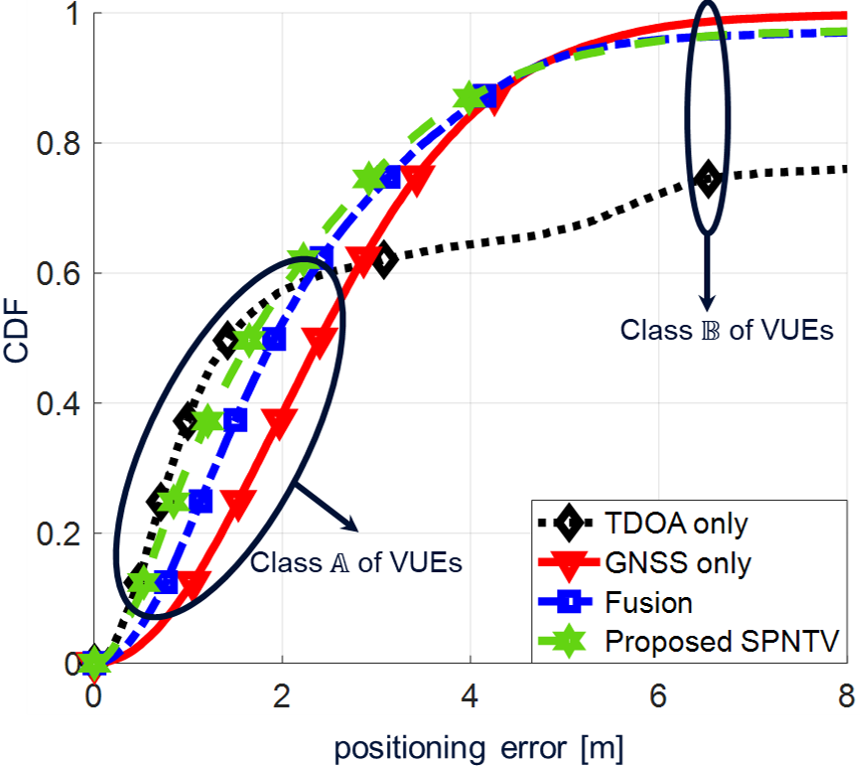}
  \caption{MAP-based positioning accuracy.}\label{fig_map}
  \vspace{-.25in}
  \end{center}
\end{figure}

In this use case, V2X UEs are randomly distributed in a six-lane urban road and a two-tier cellular network following the simulation settings in Table 1. We assume that no RSUs are deployed in the evaluations of this use case. Alternatively, V2X UEs use their connections with the macro gNBs for the DL-TDOA positioning. Fig.~\ref{fig_map} shows that the positioning error of the DL-TDOA method outperforms that of the GNSS for almost 60\% of the V2X UEs. However, the GNSS method significantly outperforms the DL-TDOA when V2X UEs suffer from a poor cellular performance. In addition, Fig.~\ref{fig_map} shows that the positioning accuracy of fusion algorithm outperforms that of the GNSS by almost .3 m for at least 55\% of V2X UEs and maintains a decent tail behavior when class $\mathbb{B}$ V2X UEs suffer from poor DL-TDOA performance. Fig.~\ref{fig_map} also reveals that the proposed SPNTV algorithm defines the best positioning technology to be used based on the V2X UE location and maintains a long-term high accuracy for both classes of V2X UEs.

\begin{figure}
  \begin{center}
  \includegraphics[width=8.5cm,height=8.5cm,,keepaspectratio]{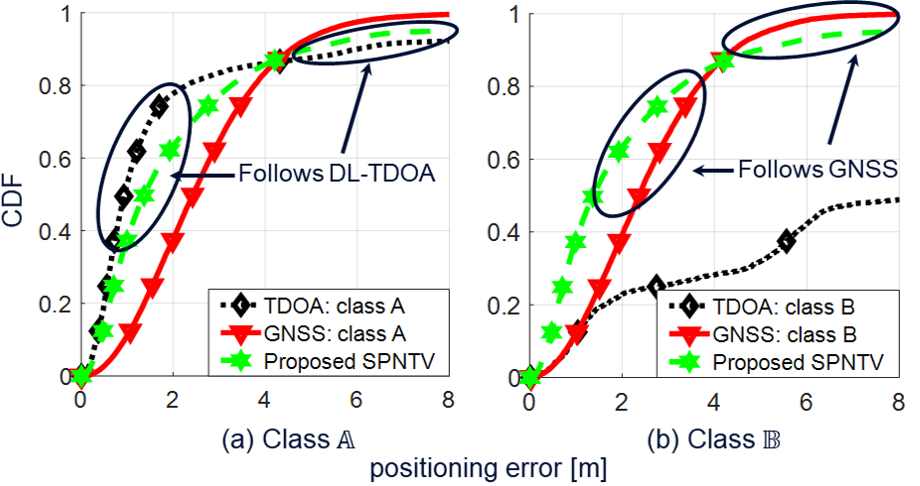}
  \caption{Positioning accuracy of two classes of V2X UEs.}\label{fig_classes}
  \vspace{-.25in}
  \end{center}
\end{figure}

In particular, the positioning accuracy of the proposed SPNTV scheme outperforms that of the GNSS and the fusion method by almost .6 m and .3 m, respectively. Fig.~\ref{fig_map} also shows that the proposed SPNTV and fusion schemes achieve a positioning error $\leq 3$ m with 76\% and 74\% availability compared to 65\% and 61\% availability when the GNSS and DL-TDOA measurements are solely used, respectively. It is worth noting that the above findings are the same if another estimator (e.g. least squares) is used for positioning. However, we use the MAP estimator (in all of the above methods) given that the positioning performance of the MAP estimator outperforms that of any other estimator that has been presented in the literature. Our numerical analysis reveals that the positioning performance of the proposed SPNTV follows that of the DL -TDOA for the class $\mathbb{A}$ of V2X UEs  (i.e., when the accuracy of DL-TDOA is $\leq \epsilon$), as shown in Fig.~\ref{fig_classes}-a. On the other hand, Fig.~\ref{fig_classes}-b shows that the SPNTV decides to use the GNSS measurements when the DL-TDOA performance significantly deteriorates. In other words, SPNTV method chooses the best positioning technology based on the location of V2X UEs. 


\begin{figure}
  \begin{center}
  \includegraphics[width=6.5cm,height=6.5cm,,keepaspectratio]{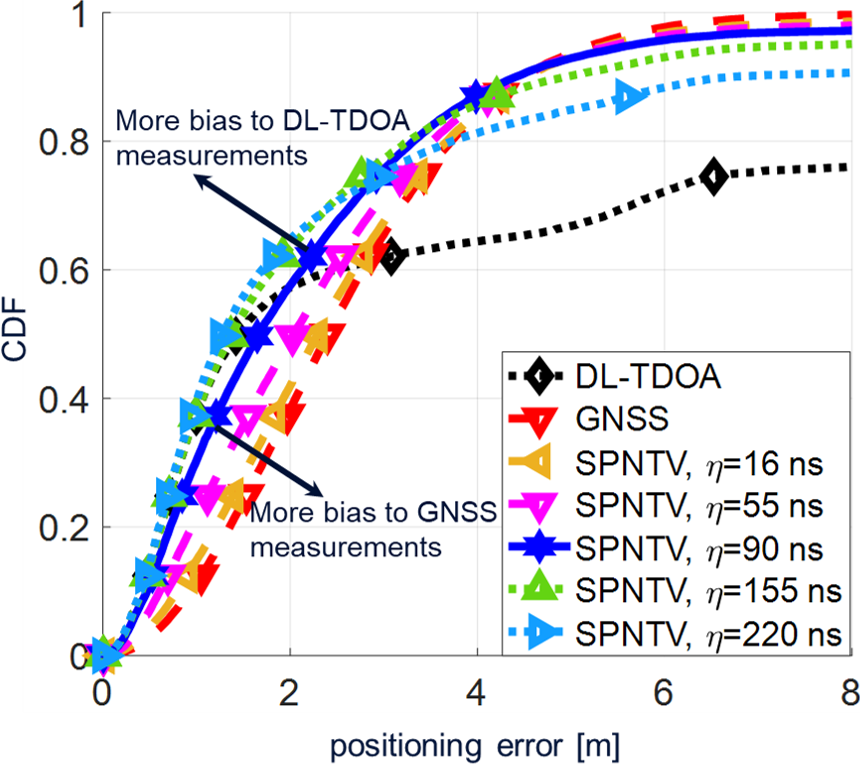}
  \caption{Fine tuning the parameter $\eta$ for the proposed SPNTV.}\label{fig_opt}
  \vspace{-.25in}
  \end{center}
\end{figure}

Fig.~\ref{fig_opt} shows the advantage of having self-inherited flexibility in the proposed SPNTV algorithm to choose the best positioning technology. In that, the selection threshold $\eta$ is fine-tuned to have more bias toward a specific technology over the other. Essentially, decreasing $\eta$ adds more bias toward using the GNSS measurements for positioning. As shown in Fig.~\ref{fig_opt}, the positioning accuracy of the proposed SPNTV algorithm mimics that of the GNSS when $\eta=16$ ns. In other words, the majority of V2X UEs decide to use their GNSS measurements for positioning. Similarly, most of V2X UEs use the cellular measurements when $\eta= 220$ ns. In contrast, it is a hard optimization problem to find the optimal weights in the MAP-based fusion algorithm, given the mutual dependency between the accuracy of the fused measurements and the deployment environment. Generally, fusing two groups of accurate and poor measurements will not lead to an improved accuracy unless there is a specific bias toward one group over the other. Hence, having a selection criterion (which is represented by $\eta$ in the proposed SPNTV) to switch between different technologies is considered as an instrumental factor to improve the positioning accuracy of any hybridized positioning algorithm for practical deployments.


\subsection{Tunnel use case}\label{case2}
\begin{figure}
  \begin{center}
  \includegraphics[width=8.5cm,height=8.5cm,,keepaspectratio]{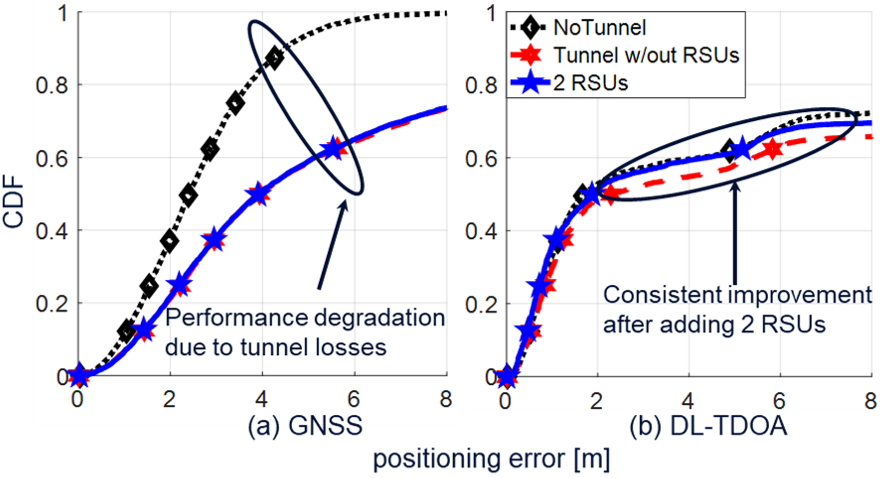}
  \caption{Performance of single-technology positioning method.}\label{fig_tunnel}
  \vspace{-.25in}
  \end{center}
\end{figure}
 In this use case, a tunnel section is added to the urban road such that V2X UEs are divided evenly into two groups, one inside the tunnel and another one outside the tunnel. Essentially, V2X UEs suffer from weak satellite and cellular signal coverage inside the tunnel due to the penetration loss and the lack of LOS connections. Thus, providing V2X UEs with a long-term high positioning accuracy is one of the main challenges in such scenarios. It is worth mentioning that the emergency brake warning is one of the typical traffic safety use cases in V2X tunnel scenarios. In that, a V2X UE is expected to apply a brake while approaching/leaving the tunnel. Hence, it is of paramount importance that V2X UEs can achieve high positioning accuracy anytime and anywhere in these scenarios. It is also clear that relying on GNSS solutions alone will not be sufficient. We utilize the proposed hybridized positioning algorithms in Section~\ref{sec_hymethods} and evaluate the positioning accuracy in this use case. Further, we investigate the potential gains of adding RSUs inside the tunnel to provide the V2X UEs with a long-term high positioning accuracy. In that, V2X UEs use their connections with the gNBs and RSUs for DL-TDOA measurements outside and inside the tunnel, respectively. 
 
\begin{figure}
  \begin{center}
  \includegraphics[width=6.5cm,height=6.5cm,,keepaspectratio]{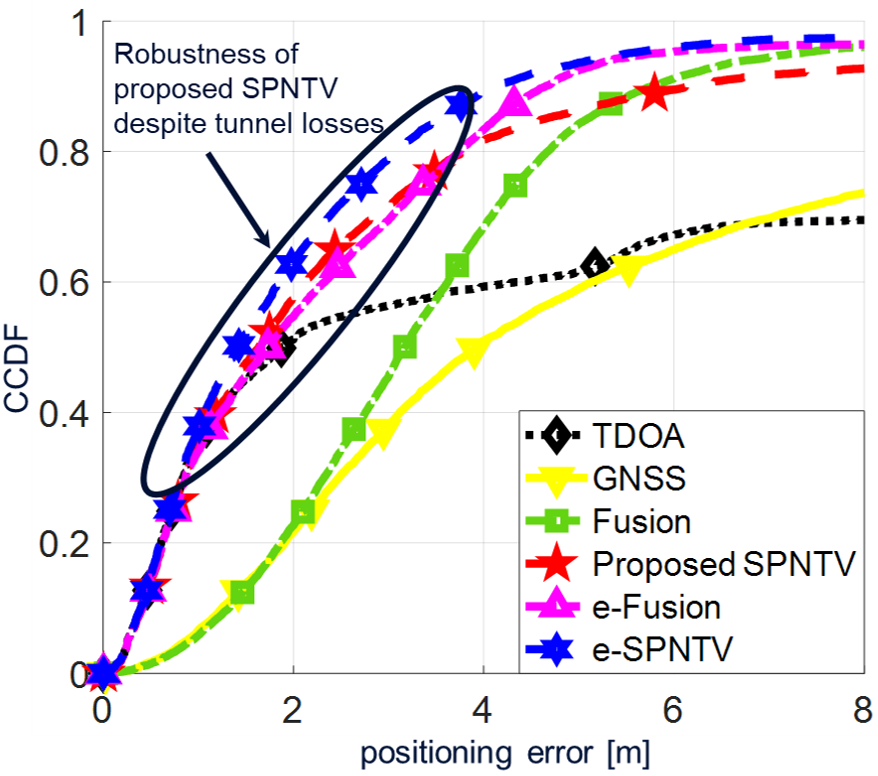}
  \caption{Overall positioning error in the tunnel use case.}\label{fig_err2}
  \vspace{-.25in}
  \end{center}
\end{figure}

Fig.~\ref{fig_tunnel}-a shows a severe performance degradation in GNSS positioning at almost 80\% of V2X UEs due to the tunnel losses. On the other hand, Fig.~\ref{fig_tunnel}-b shows that the performance degradation in cellular positioning is smaller than that of the GNSS. It also shows that adding two RSUs (at the tunnel edges) improves the DL-TDOA performance to mimic that of the baseline scenario, i.e., urban road use case without a tunnel. In this regard, Fig.~\ref{fig_err2} demonstrates the robustness of the proposed SPNTV method against tunnel losses. Essentially, the proposed SPNTV ensures providing V2X UEs with high positioning accuracy on a continuous-time basis as they move in and out of the tunnel. It also shows that the fusion algorithm suffers from a performance degradation of almost 2 m for more than 85\% of V2X UEs compared with the urban road use case (see Fig.~\ref{fig_map}) due to the tunnel losses. In other words, the fusion algorithm is more susceptible to any accuracy degradation in either one of the positioning technologies (cellular or GNSS). This is because it is hard to fine-tune the fusion method to have more bias to one technology over the other. On the other hand, Fig.~\ref{fig_err2} reveals that, with a proper configuration of the SPNTV algorithm, it is possible to maintain high accuracy for almost 80\% of V2X UEs despite the tunnel losses (see Fig.~\ref{fig_map} for comparison with the SPNTV performance in an urban road use case). This can be achieved by adding more bias toward the cellular measurements, i.e., $\eta = 125$ ns. 

It is worth noting that a significant increase of the bias toward the cellular measurements will result in degrading the SPNTV performance to mimic that of the DL-TDOA. To solve this problem, we evaluate the performance of enhanced versions of the hybridized positioning schemes presented in Section~\ref{sec_hymethods}. In that, V2X UEs use the DL-TDOA method inside the tunnel and either the SPNTV or the fusion algorithm outside the tunnel. As shown in Fig.~\ref{fig_err2}, the positioning accuracy of the enhanced schemes outperforms using either the SPNTV or the fusion algorithm all the time. In particular, e-SPNTV, e-fusion achieve a positioning error $\leq 3$ m with 80\% and 70\% availability compared to 71\%, 56\%, 45\%, and 38\% availability when SPNTV, DL-TDOA, fusion, and GNSS are used, respectively. Essentially, the e-SPNTV method allows V2X UEs to dynamically activate and deactivate the GNSS measurements based on their location. This can be interpreted as another degree of freedom, in which, the selection parameter $\eta$ is dynamically configured based on the V2X UE location rather than using a single static value all the time. Further, the e-fusion method allows V2X UEs to overcome the inevitable GNSS performance degradation due to the tunnel losses by using only the cellular measurements inside the tunnel. 

\begin{figure}
  \begin{center}
  \includegraphics[width=8cm,height=8cm,,keepaspectratio]{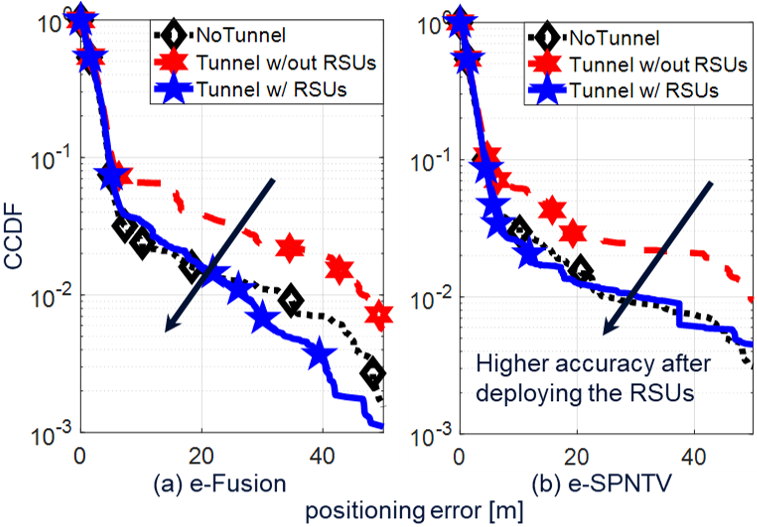}
  \caption{Positioning performance with the RSUs.}\label{fig_rsus}
  \vspace{-.25in}
  \end{center}
\end{figure}

Finally, the findings of Fig.~\ref{fig_err2} are the same when the RSUs are not deployed at the tunnel edges except with slight performance degradation due to the DL-TDOA performance degradation. In this regard, we investigate the performance of the most vulnerable V2X UEs in three scenarios (1) urban road without a tunnel, (2) tunnel without RSUs, and (3) tunnel with RSUs using the e-Fusion and e-SPNTV methods. Fig.~\ref{fig_rsus} shows that deploying just two RSUs at the tunnel edges significantly improves the tail of the UE's positioning error distribution, i.e., the positioning performance of the most susceptible V2X UEs to the high positioning errors, by more than $\approx$ 56\%.

\section{Conclusion}\label{sec_conc}
In this paper, we investigate the potential solutions to provide the NR V2X UEs with long-term and high-precision positioning in the future 3GPP releases. Specifically, we propose a novel selection positioning method to dynamically switch between the GNSS and DL-TDOA measurements based on the locations of V2X UEs and the accuracy of the collected measurements. Further, we utilize a MAP-based estimator to evaluate the performance gains of fusing the measurements from different technologies in NR V2X services. The proposed hybridized methods are evaluated via extensive system-level simulations which demonstrate that the proposed algorithms can achieve a positioning error $\leq$ 3 m with $\approx $ 76\% availability compared to $\approx $ 51\% and 58\% availability when GNSS and DL-TDOA measurements are solely used, respectively. Further, we leverage the deployment of the RSUs to improve the tail of the UE's positionig error distribution by more than $\approx$ 56\%.

\balance
\bibliographystyle{IEEEtran}
\bibliography{IEEEabrv.bib,Bibliography.bib}
\end{document}